\newcommand{\mn}{{\mu\nu}}
\newcommand{\Tmn}{T_{\mu\nu}}
\newcommand{\TRmn}{T^{(R)}_{\mu\nu}}
\newcommand{\rs}{{\rho\sigma}}
\newcommand{\Trs}{T_{\rho\sigma}}
\newcommand{\TRrs}{T^{(R)}_{\rho\sigma}}
\newcommand{\mnrs}{{\mu\nu\rho\sigma}}

\newcommand{\trphitwo}{\Tr \phi^2}

\documentclass{PoS}

\title{Towards a holographic description of cosmology:\\ 
Renormalisation of the energy-momentum tensor of the dual QFT}

\ShortTitle{EMT renormalisation for holographic cosmology}

\author{\speaker{Joseph K. L. Lee}$^1$, Luigi Del Debbio$^1$, Andreas J\"{u}ttner$^2$, Antonin Portelli$^1$, Kostas Skenderis$^3$\\
        $^1$School of Physics and Astronomy, University of Edinburgh, Edinburgh EH9 3JZ, UK\\
        $^2$STAG Research Centre and School of Physics and Astronomy, University of Southampton, Southampton SO17 1BJ, UK\\
        $^3$STAG Research Centre and Mathematical Sciences, University of Southampton, Southampton SO17 1BJ, UK\\
        E-mail: \email{joseph.lee@ed.ac.uk}}

\author{LatCos Collaboration}

\abstract{In the holographic approach to cosmology, cosmological observables are described in terms of correlators of a three-dimensional boundary quantum field theory. As a concrete model, we study the 3$d$ massless $SU(N)$ scalar matrix field theory. In this work, we focus on the renormalisation of the energy-momentum tensor 2-point function, which can be related to the CMB power spectra. Here we present a non-perturbative procedure to remove divergences resulting from the loss of translational invariance on the lattice, by imposing Ward identities. This will allow us to make predictions for the CMB power spectra in the regime where the dual QFT is non-perturbative.}

\FullConference{37th International Symposium on Lattice Field Theory - Lattice2019\\
		16-22 June 2019\\
		Wuhan, China}

\begin{document}

\section{Introduction}
Holographic cosmology is a new framework for describing the very early Universe, the period usually associated with inflation. The holographic approach to cosmology allows us to describe cosmological observables in terms of correlators of a dual three-dimensional quantum field theory (QFT) with no gravity. This project aims to compute non-perturbatively the energy-momentum tensor (EMT) 2-point function of the dual QFT on the lattice, which enables us to obtain the predictions for the cosmic microwave background (CMB) power spectra. On the lattice, the EMT renormalises due to the breaking of continuous translational symmetry. This work proposes a method to renormalise the EMT of the 3$d$ massless $SU(N)$ scalar matrix field theory using Ward identities, which allows us to compare the CMB power spectrum prediction against Planck data.

\section{Holographic Cosmology}
The CMB is a unique window into the physics of the very early Universe. The standard model of cosmology, the so-called $\Lambda$CDM model, provides an excellent fit to the CMB primordial power spectrum ($\Delta^2_{\mathcal{R}}$) observed, e.g. from Planck \cite{planck}, via a power law parametrisation. This near-power-law spectrum is usually explained by and considered a success of the theory of cosmic inflation, where gravity coupled to matter is perturbatively quantized around an accelerating FRW background. However, inflation is only an effective theory and does not describe what happens at earlier times, when the spacetime curvature is sufficiently high, quantum gravity effects are expected to be significant, and the usual geometric picture with metric perturbation is expected to break down.

Holographic cosmology is a new framework for cosmology that utilizes the holographic principle, which maps gravitational dynamics of a $(d+1)$ dimensional theory to observables of a $d$ dimensional QFT with no gravity. In the context of early-times cosmology, it maps cosmological observables such as the CMB power spectra to correlation functions of the EMT of the dual QFT. A class of qualitatively new (non-geometric) models for the very early Universe has been put forward in \cite{mcfadden1}, in which the dual QFT is a super-renormalisable QFT containing massless fields in the adjoint of $SU(N)$.

The main cosmological observables to be tested are the CMB primordial power spectra. They are related to the momentum space 2-point function of the EMT of the dual QFT $\Tmn$,
\begin{equation}
    \llangle \Tmn (q) \Trs (-q) \rrangle = A(q) \Pi_\mnrs + B(q) \pi_\mn \pi_\rs,
\end{equation}
where $\expval{\Tmn (q_1) \Trs (q_2)} = (2\pi)^3 \delta^3(q_1+q_2)\llangle \Tmn (q) \Trs (-q)\rrangle$, $\pi_\mn = \delta_\mn - \frac{q_\mu q_\nu}{q^2} $ is a transverse projector and $\Pi_\mnrs = \frac{1}{2}\qty(\pi_{\mu\rho}\pi_{\nu\sigma}+\pi_{\mu\sigma}\pi_{\nu\rho}-\pi_\mn \pi_\rs)$ a transverse-traceless projector. The explicit holographic formulae for the power spectra are given by
\begin{equation}
\label{eq:power_spectra_AB}
    \Delta^2_{\mathcal{R}} = - \frac{q^3}{16 \pi^2} \frac{1}{B(q)} \text{\quad and \quad} \Delta^2_{T} = - \frac{2q^3}{\pi^2} \frac{1}{A(q)}.
\end{equation}
For the class of theories in \cite{mcfadden1}, perturbation theory up to 2 loops gives the scalar power spectrum
\begin{equation} \label{eq:power_spectrum_hc}
    \Delta^2_{\mathcal{R}-\text{HC}}(q) = \frac{\Delta_0^2}{1+(gq^*/q) \ln \abs{q/\beta g q^*}}.
\end{equation}
Here $q$ is the comoving wave number, $q^*$ an arbitrary pivot scale, $\beta$, $g$, and $\Delta_0^2$ parameters which depend on the field content of the dual QFT. This is distinct from the $\Lambda$CDM parametrisation:
\begin{equation}
    \Delta^2_{\mathcal{R}-\Lambda \text{CDM}}(q)=\Delta_0^2 \left( \frac{q}{q^*}\right)^{n_s-1},
\end{equation}
where $n_s$ the spectral index. 

The predicted power spectrum \eqref{eq:power_spectrum_hc} has been tested against WMAP and Planck data in \cite{easther, afshordi1, afshordi2}. From the fit with Planck data, it was found that the holographic description (restricted to the region where perturbation theory is valid) was competitive with the $\Lambda$CDM model; the data is consistent with the dual QFT being a Yang-Mills (YM) theory coupled to non-minimal scalars with quartic interactions (with no fermions; the data rule out the dual theory being YM coupled to fermions only).

The class of massless super-renormalisable QFT under consideration, with a coupling $g_\text{YM}^2$ with mass dimension 1, suffers from severe infrared (IR) divergences in perturbation theory. Correlation functions are expanded perturbatively in powers of the effective dimensionless coupling $g_\text{eff} \equiv g_\text{YM}^2 N/q$. For each increasing order in perturbation theory, the IR degree of divergence increases by one. At low momenta ($g^{-1}_\text{eff} < 1$), IR divergences appear and the introduction of an arbitrary infrared cutoff is necessary. It is believed that these IR divergences are only artefacts of perturbation theory, and the full theory is non-perturbatively IR finite (whereby the coupling $g_\text{YM}^2$ effectively acts as an IR regulator) \cite{jackiw, appelquist}. Furthermore, from the comparison with Planck data, the dual theory is strongly coupled in the low-multipole region of the CMB spectrum ($\ell < 30$). This means that perturbative calculation cannot succeed in describing this part of the spectrum. 

\section{Lattice Simulation}
The above issues can be solved by performing computations in the dual theory using non-perturbative lattice simulations. The aim of the project is to evaluate the EMT 2-point function non-perturbatively on the lattice in order to reliably test the predictions of these holographic models.
\subsection{The Model}
The model we have simulated is a 3$d$ scalar field theory with $\phi$ in the adjoint of $SU(N)$ (i.e., $\phi$ is an $N \times N$ traceless Hermitian matrix) and a $\phi^4$ interaction. This is a simplified version of the preferred theory with gauge interactions. The coupling constant multiplying the $\phi^4$ term plays the role of the Yang-Mills coupling and one may rescale the scalar field such that it appears as an overall constant in the action. Upon discretization the model is governed by the action
\begin{equation}
    S[\phi]=\frac{a^3N}{g}\sum_{x\in \Lambda^3}\Tr \qty{\sum_\mu \qty[\delta_\mu  \phi (x) ]^2+m^2\phi(x)^2+\phi(x)^4}.
\end{equation}
Here $\delta_\mu$ is the forward finite difference operator defined by $\delta_\mu \phi(x) = a^{-1}[\phi(x+a\hat{\mu})-\phi(x)]$, where $\hat{\mu}$ is the unit vector in direction $\mu$, $\Lambda$ a lattice with cubic geometry containing $N_L^3$ points, $a$ the lattice spacing, $g = g_\text{YM}^2 N$ the 't Hooft coupling constant (with dimension 1), and $m$ the bare mass of the theory, which renormalises additively. We are interested in the massless limit, and the massless point of the theory is determined by evaluating the Binder cumulant using the method described in \cite{juettner}. We have so far performed simulations for $N$ up to 3, containing ensembles with 4 lattice spacings and 2 masses close to the massless point.

\subsection{Non-Perturbative Window}

\begin{figure}
    \centering
    \includegraphics[width=.6\textwidth]{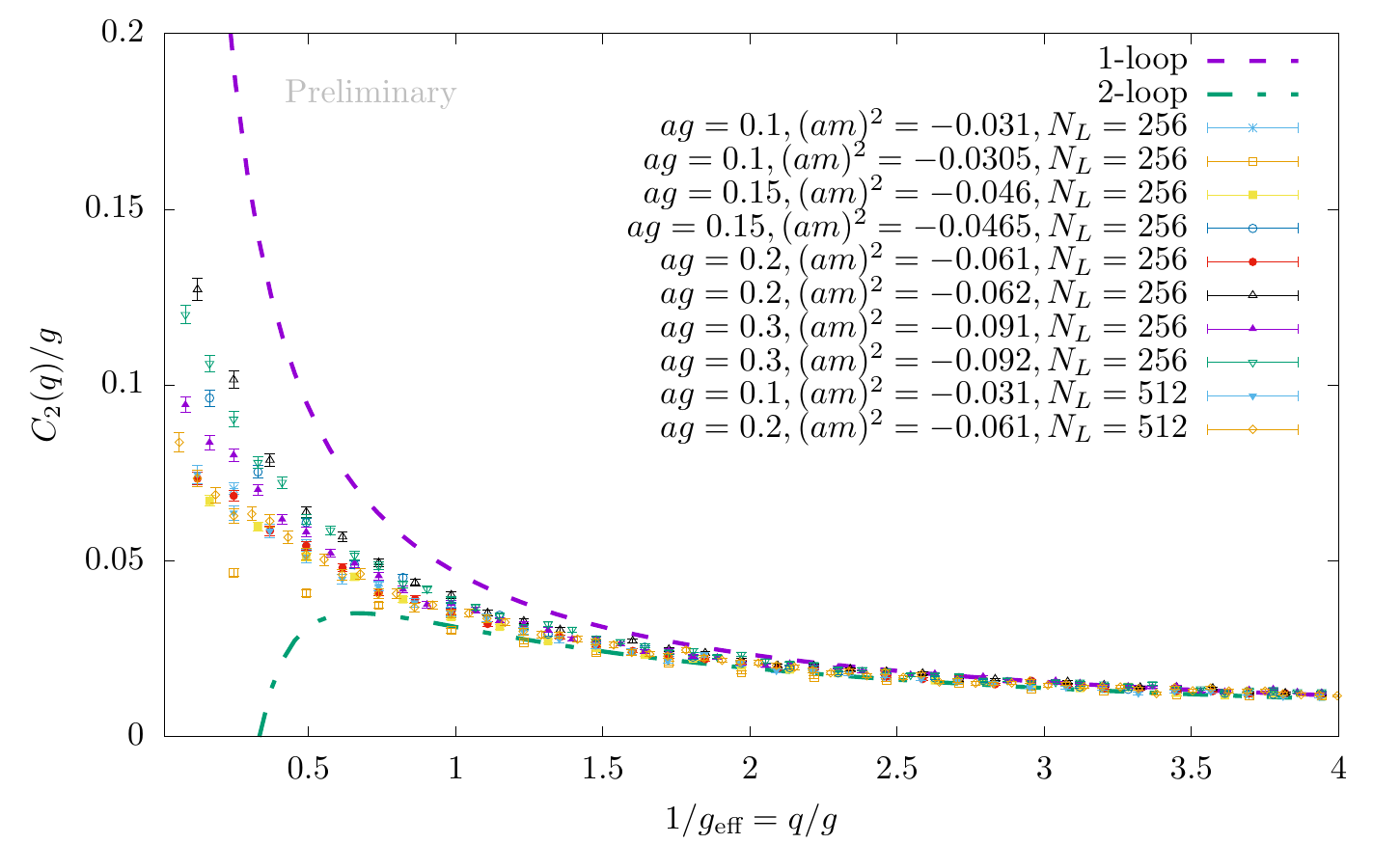}
    \caption{Plot of the 2-point function $C_2 \qty(q=(0,0,q_2))$ of $\Tr \phi^2$ in equation \eqref{eq:c2}. The dashed lines are the perturbative predictions at the 1-loop and 2-loop order in lattice perturbation theory. The lattice data come from $N=2$ simulations at different lattice spacings and masses near the massless point.}
    \label{trphi2trphi2}
\end{figure}

As an illustration of the IR finiteness of the non-perturbative simulations, the 2-point function of the operator $\trphitwo$,
\begin{equation} \label{eq:c2}
    C_2(q) = a^3 \sum_{x \in \Lambda}\ e^{-iq\cdot x} \expval{\trphitwo (x) \trphitwo (0)},
\end{equation}
is plotted in figure \ref{trphi2trphi2}. In the perturbative region ($g^{-1}_\text{eff}>1$), the lattice simulation is very well described by the results from perturbation theory; however, in the non-perturbative window ($g^{-1}_\text{eff}<1$), rather than diverging as predicted by the perturbative expansion, our lattice results suggest that it is in fact IR finite, as expected for our super-renormalisable theory. 

\section{Energy-Momentum Tensor Renormalisation}

For the $SU(N)$ scalar matrix theory, the na\"{i}ve discretisation of the EMT is
\begin{equation}
    \Tmn = \frac{N}{g} \Tr\qty{2\qty(\bar{\delta}_\mu \phi)\qty(\bar{\delta}_\nu \phi) - \delta_\mn\qty[\sum_\rho\qty(\bar{\delta}_\rho \phi)^2+m^2 \phi^2 + \phi^4)]},
\end{equation}
where the central finite difference $\bar{\delta}_\mu \phi (x) = \frac{1}{2a}[\phi(x+a\hat{\mu})-\phi(x-a\hat{\mu})]$ is used to obtain a Hermitian operator.

In the continuum, due to translational invariance, the EMT satisfies Ward identities (WI) of the form
\begin{equation}
\label{eq:cont_ward}
    \expval{\partial^\mu \Tmn(x) P(x_i)} = -\expval{\frac{\delta P(x_i)}{\delta\phi (x)} \partial_\nu \phi (x)}.
\end{equation}
For operators $P$ which are ultraviolet (UV) finite, the right hand side of equation \eqref{eq:cont_ward} is finite at separated points and correlation functions of the divergence of the EMT on the left hand side are also finite up to contact terms. For this theory, it can be further shown that the insertion of $\Tmn$ does not introduce new UV divergences.

On the lattice, the continuous translation group is broken into the discrete subgroup and $\Tmn$ is no longer conserved. Now, the WI on the lattice has to include an additional term,
\begin{equation}
    \expval{\bar{\delta}^\mu \Tmn(x)P(x_i)} = -\expval{\frac{\delta P(x_i)}{\delta\phi (x)} \partial_\nu \phi (x)} + \expval{X_\nu (x)P(x_i)},
\end{equation}
where $X_\nu$ is an operator proportional to $a^2$, which vanishes in the continuum limit. However, radiative correction causes the expectation value $\expval{X_\nu(x)P(x_i)}$ to produce a linearly $a^{-1}$ divergent contribution to the WI. Therefore, the na\"{i}vely discretized EMT will not render the WI finite when the regulator is removed; $\Tmn$ has to be renormalised by subtracting the divergent operator mixing.

In four dimensions, it has been shown in \cite{caracciolo} using symmetry and dimension counting that $\Tmn$ potentially mixes with 5 lower-dimensional operators, generating these divergences. However, in three dimensions, mixing only occurs with $O_3 = \delta_\mn \trphitwo$. The renormalised EMT on the lattice can therefore be expressed as
\begin{equation}
    \Tmn^{(R)} =\Tmn - \frac{c_3}{a} \delta_\mn \trphitwo.
\end{equation}
where $c_3$ is the coefficient correcting for the divergent operator mixing on the lattice, which has to be tuned to satisfy the WI \eqref{eq:cont_ward} up to discretisation effects, so that the continuum finite WI is recovered in the continuum limit.

\subsection{Renormalisation Condition}
To obtain the operator mixing, we consider the following momentum space 2-point function,
\begin{equation}
    C_\mn (q) = a^3 \sum_{x \in \Lambda}\ e^{-iq\cdot x} \expval{\TRmn (x) \trphitwo (0)}.
\end{equation}
In the infinite volume limit, the connected part of the 2-point function vanishes at $q=0$ as a consequence of the WI and $O(3)$ invariance (assuming that the theory is IR finite); however, with finite lattice spacing, a constant term $C_\mn(0)$ appears, which can be subtracted
\begin{equation}
    \hat{C}_\mn (q) = C_\mn(q) - C_\mn(0).
\end{equation}
We impose as a renormalisation condition the Ward identity to $\hat{C}_\mn(q)$, which in momentum space is
\begin{equation}
    \bar{q}_\mu \hat{C}_\mn(q) = 0,
\end{equation}
where $\bar{q}=\frac{1}{a}\sin \qty(aq)$ is the lattice momentum. This implies that $\hat{C}_\mn (q)$ is transverse, i.e.,
\begin{equation}
    \hat{C}_\mn(q)  = F(q) \bar{\pi}_\mn,
\end{equation}
where $\bar{\pi}_\mn$ is defined as the transverse projector with lattice momentum $\bar{q}$. To apply this condition to lattice data, we require the correlator to vanish with momentum purely in the transverse direction, e.g.,
\begin{equation} \label{eq:renorm_condition}
    \hat{C}_{22}\qty(q=\qty(0,0,q_2))=0.
\end{equation}

The value of $c_3$ can be evaluated in lattice perturbation theory, which up to 1 loop is equal to $0.1$ for $N=2$. We are currently exploring the precise implementation of the renormalisation condition \eqref{eq:renorm_condition} to the lattice data, particularly the systematic errors associated; our preliminary analysis suggests that it is consistent with the perturbative calculation.

\subsection{2-Point Function of the Energy-Momentum Tensor}
\begin{figure}
    \centering
    \includegraphics[width=.6\textwidth]{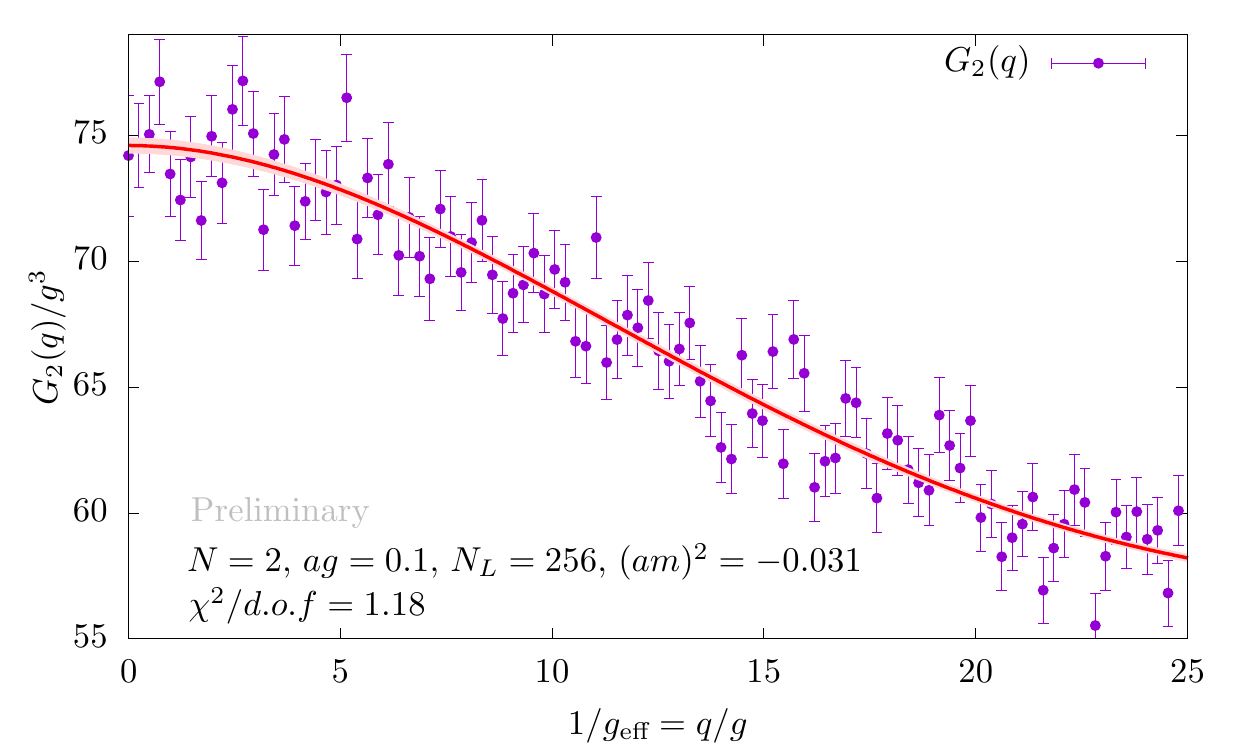}
    \caption{The form factor $G_2(q)= C_{0101}\qty(q=(0,0,q_2))$ with a correlated fit using equation \eqref{eq:G-form-factor}}
    \label{t01t01}
\end{figure}

Using the renormalised EMT, we evaluate the EMT 2-point function 
\begin{equation} \label{eq:4comp}
    C_{\mnrs} (q) = a^3 \sum_{x \in \Lambda}\ e^{-iq\cdot x}\ \expval{\TRmn (x) \TRrs (0)}
\end{equation}
from the lattice simulation. For later convenience, we decompose the correlator into form factors
\begin{equation}
\begin{split}
    C_{\mnrs} (q) = &G_1(q)\delta_{\mu\nu}\delta_{\rho\sigma} \\
        +&G_2(q) (\delta_{\mu\rho}\delta_{\nu\sigma} + \delta_{\mu\sigma}\delta_{\nu\rho}) \\
        +&G_3(q) (\delta_{\mu\nu}q_\rho q_\sigma + \delta_{\rho\sigma}q_\mu q_\nu)/q^2\\
        +&G_4(q)(\delta_{\mu\rho}q_\nu q_\sigma + \delta_{\mu\sigma}q_\nu q_\rho + \delta_{\nu\rho}q_\mu q_\sigma + \delta_{\nu\sigma}q_\mu q_\rho)/q^2  \\
        +&G_5(q) q_\mu q_\nu q_\rho q_\sigma/(q^2)^2.
\end{split}
\end{equation}
Here we assumed $O(3)$ covariance; there are in fact more terms present which are covariant under the cubic group on the lattice. The form factors $G_i(q)$ can be evaluated by choosing suitable combinations of indices and momenta from the lattice data; for example, $G_2(q)$ can be obtained from $C_{0101}\qty(q=(0,0,q_2))$, which is shown in figure \ref{t01t01}. Our lattice data suggests a good fit for the form factors $G_i(q)$ using the following function:
\begin{equation}
\label{eq:G-form-factor}
    G_i(q) = \frac{\alpha_i}{a^3}+\frac{\beta_i}{a} \hat{q}^2 + \gamma_i \hat{q}^3,
\end{equation}
where $\hat{q}=\frac{2}{a}\sin \qty(\frac{aq}{2})$ is the lattice momentum, $\beta_i$ and $\gamma_i$ dimensionless coefficients. Each form factor $G_i(q)$ still contain UV divergences proportional to $\frac{1}{a^3}$ and $\frac{q^2}{a}$, which comes from the $x \to 0$ limit in \eqref{eq:4comp} even though the operators $\Tmn$ are renormalised. We are currently exploring the method to subtract these divergences to obtain the renormalised form factors, which will in principle allow us to reconstruct $A(q)$ and $B(q)$ from equation \eqref{eq:power_spectra_AB} to obtain the power spectra predictions.

\section{Conclusion \& Outlook}
In this work we have proposed a scheme to renormalise the EMT 2-point function using Ward identities, and discussed the method to obtain form factors by fitting our lattice data, which will allow us to test the holographic description of the very early Universe using non-perturbative lattice simulations. In the future we will extrapolate our results to the massless, infinite volume, continuum limit, then compare our lattice results against Planck CMB data. We will then repeat the analysis for theories containing scalar fields coupled to gauge fields, which is the theory preferred from fitting the perturbative results with Planck data.

\acknowledgments
A.P. and J.K.L.L. are funded in part by the European Research Council (ERC) under the European Union's Horizon 2020 research and innovation programme under grant agreement No 757646. J.K.L.L. is also partly funded by the Croucher foundation through the Croucher Scholarships for Doctoral Study. A.J. and K.S. received funding from STFC consolidated grant ST/P000711/1. L.D.D. is supported by an STFC Consolidated Grant, ST/P0000630/1, and a Royal Society Wolfson Research Merit Award, WM140078. Lattice computations presented in this work has been performed using the Cambridge Service for Data Driven Discovery (CSD3), part of which is operated by the University of Cambridge Research Computing on behalf of the STFC DiRAC HPC National e-Infrastructure.

\end{document}